# Superellipsoid modeling of wire-laser additive manufacturing defects measured by digital image correlation

Khalil Hachem[1], Yann Quinsat[1], Christophe Tournier[1], Nicolas Beraud[2]

[1]*Université Paris-Saclay, ENS Paris-Saclay, LURPA, Gif-sur-Yvette, France, 91190*
[2]*Univ. Grenoble Alpes, CNRS, Grenoble INP, G-SCOP, Grenoble, France, 38000*

*khalil.hachem@ens-paris-saclay.fr*

**Abstract**

In wire-laser additive manufacturing (WLAM), producing parts with multiple entities requires optimizing both additive and subtractive processes. This involves adjusting positions and integrating machining operations. Geometric defects that occur during the components' deposition can affect subsequent entity production, making real-time measurement crucial. A contactless measuring system is essential to assess geometric deviations without complex post-processing. A comprehensive image stereocorrelation approach addresses this challenge. The produced geometry often has unaccounted features, such as rounded edges and corners. Superellipsoids serve as a versatile modeling tool for representing common geometries in WLAM and assessing dimensional and shape defects. Defined by three scale and two shape parameters, superellipsoids are introduced generally. This modeling tool is then combined with a global image correlation problem. The parameters defining this geometry are identified by minimizing the gray level difference between two images of a part. This approach is demonstrated in a case study, highlighting deviations in WLAM with a parameterized surface and paving the way for geometry updates in computer-aided manufacturing.

Wire-Laser Additive Manufacturing, Digital image correlation, Defects modeling, Superellipsoids

**1. Introduction**

A new era in industrial manufacturing has emerged with Additive Manufacturing (AM), offering significant advantages over traditional subtractive methods by enabling the production of complex parts and repairing damaged components [1]. Wire-Laser Additive Manufacturing (WLAM) is a specific AM form where metal wire is fed to a substrate, intersecting with a concentrated laser to create a melt pool for layer-by-layer deposition. The machine comprises a laser system and an automated wire feed and is typically operated by a robotic arm.

However, defects often occur in additively manufactured parts due to various factors, including improper equipment performance, process parameters, feedstock quality, and heat accumulation. Defects can be categorized into geometrical, morphological, and microstructural anomalies [2]. This paper focuses on external geometrical defects at the entity level, not the beads. A post-processing subtractive phase is often utilized to enhance surface finish and address defects.

Accurate *in-situ* measurement of part deviations is essential for effective production, allowing for adjustments to the machining and additive processes. A contactless measuring system is necessary for evaluating geometrical defects without disrupting production. Optical signal detection using CMOS cameras provides valuable full-field measurements for assessing part deformation [1]. Techniques like 3D Digital Image Correlation (DIC) have been used effectively for *in-situ* measurements, improving production quality control [3].

Superellispoids (SE) are flexible mathematical tools introduced by Alan Barr in 1981 [4]. They represent a range of shapes, from simple geometric figures to complex forms. SE simplify shape manipulation and transformation, making them crucial in computer graphics for object and surface modeling and computer vision and robotics for object representation.

This contribution uses the above geometrical representation to model the shape and dimensional defects of parts produced by the WLAM process at the entity level. Given this perspective, this approach couples the geometrical modeling using SE with the 3D-DIC measuring technique. It also highlights the advantage of such implicit forms in image correlation optimization. Consequently, a case study is proposed to retrieve a SE representing the actual geometry of a produced part. This paper is structured as follows: the superellipsoids and their generic geometries are defined in Sec. 2. Then, the correlation between the measuring technique and the modeling approach is described in Sec. 3 alongside a case study. Section 4 presents the results, concludes the advantages of the geometrical model, and presents the use of this developed approach for future work.

**2. Geometrical modeling: Superellipsoids**

A superellipsoid is a three-dimensional shape formed by taking the spherical product of two two-dimensional models, also known as superellipses. Only five parameters define it. The 3D vector $\mathbf{X}_{SE}$ representing the SE surface is stated in *Eq. 1*.

$$\mathbf{X}_{SE}(u,v) = \begin{bmatrix} a_x \cos^{\epsilon_1}(u) \cos^{\epsilon_2}(v) \\ a_y \cos^{\epsilon_1}(u) \sin^{\epsilon_2}(v) \\ a_z \sin^{\epsilon_1}(u) \end{bmatrix} \quad (Eq.1)$$

$a_x$, $a_y$, and $a_z$ are scale parameters that control the superellipsoid dimension along the x, y, and z axes, respectively. The parameters $\epsilon_1$ and $\epsilon_2$ control the shape curvature; they express squares along the z-axis and the xy-plane. The parametric angles u and v vary within the following limits: $-\pi/2 \leq u \leq \pi/2, -\pi \leq v \leq \pi$. The superellipsoid allows for diverse three-dimensional shapes by varying the parameters $\epsilon_1$ and $\epsilon_2$, enhancing its design versatility. It remains convex for

$0 \leq \epsilon_1, \epsilon_2 \leq 2$. *Figure 1* illustrates the different geometries achievable through these shape parameter adjustments. Cuboids, for example, are obtained when both $\epsilon_1$ and $\epsilon_2$ are less than 1. Therefore, the vertical edges can be rounded, and the sides are transformed into cylindrical bodies with an $\epsilon_2$ closer to 1. Cuboid-shaped components are easy to manufacture using AM. Moreover, complex parts can be divided into multiple entities, each modeled by one of these superellipsoids.

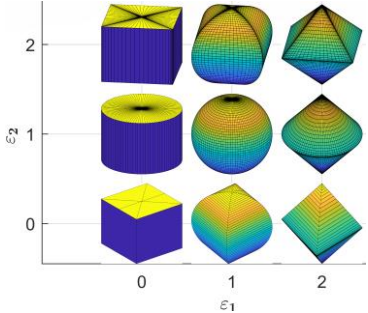

**Figure 1.** Variation of SE geometries with different shape parameters.

### 3. DIC measurements and SE modeling

This section introduces the methodology and mathematical formulation needed for modeling SE with image correlation measurement. An experimental application is then introduced to validate the developed approach and highlight the advantages of using the superellipsoid as a geometrical model.

#### *3.1. Methodology*

A pair of images of the measurand and the nominal 3D model are essential for image correlation measurement and geometric modeling. This nominal 3D set is projected into 2D in the captured images using the left and right projection matrices $[P^l], [P^r]$. The primary goal of the global image correlation process is to deform the nominal geometry to ensure grey-level conservation between the two images. Due to acquisition noise, perfect grey-level alignment is not achievable. Consequently, it is crucial to establish a global formulation of the problem, in which the sum of the least squares is represented as an image correlation residual in the parametric space. This sum is then linearized and minimized by optimizing the SE shape defined by its five key parameters $\mathbf{Y}$ (*Eq. 2*). This optimization enables the retrieval of a 3D geometry, whose 2D projection in the images results in a minimal difference in grey-level intensities, thereby improving the accuracy of the geometrical representation.

$$\{\mathbf{Y}_{opt}\} = \underset{\{\mathbf{Y}_{opt}\}}{\operatorname{argmin}} \sum_{ROI} \left( f^l\left([P^l](X)\right) - f^r\left([P^r](X)\right) \right)^2 \quad (Eq.\ 2)$$

#### *3.2. Experimental application*

This study requires *in-situ* measurements to evaluate the geometry produced by WLAM and model it with implicit forms like superellipsoids in this case. An experimental setup has been created with two high-resolution Allied Vision cameras, each having 20.4 megapixels and fixed at a $500mm$ working distance. The cameras' intrinsic parameters and relative positions are computed using a calibration target and Matlab software. Next, a geometry for the case study is selected and designed with computer-aided design software. The additive trajectories are generated using computer-aided manufacturing software in the following step. The nominal geometry is a $40mm \times 40mm \times 30mm$ cuboid corresponding to a SE with the following scale parameters $(a_x, a_y, a_z) = (20, 20, 15)$. It is also created with rounded edges having a $15mm$ radius. This geometry corresponds to a SE with $\epsilon_1 = 0$, and $0 < \epsilon_2 < 1$.

The produced part in WLAM is shown in *Fig. 2a*. A circular pattern is projected onto the part using an XGIMI Elfin 1080p projector, which offers 600 lumens (ISO) brightness. Then, *in-situ* left and right images are captured (*Fig. 2b*). These images are also used for the necessary extrinsic self-calibration of the cameras. This step minimizes the grey level difference between the set of images while optimizing the projection matrices [6].

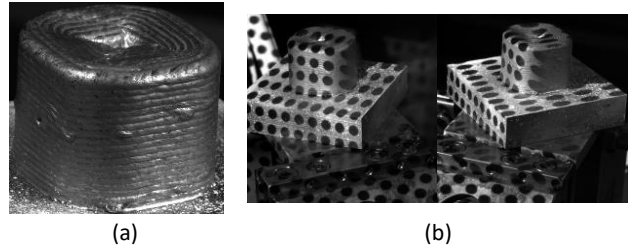

**Figure 2.** The produced WLAM part (a) and its *in-situ* images (b).

### 4. Results and Conclusions

With the *in-situ* measurements and a global DIC approach implemented, a SE is optimized to the following parameters: $(a_x, a_y, a_z, \epsilon_1, \epsilon_2) = (18.82, 18.92, 14.45, 0.04, 0.69)$. *Figure 3* illustrates rapid and smooth convergence achieved in nine iterations. This expected outcome is a key reason for choosing the geometrical model, which consists of only five parameters corresponding to five degrees of freedom. This design facilitates quick measurements and modeling during production and allows for geometry updates for future operations.

The part is digitized using the ATOS Core 3D scanner verified by VDI/VDE 2634 Part.3 testing with a $3\mu m$ maximum shape error. The optimized SE using DIC is compared to the scanned point cloud (*Fig. 3*), revealing an STD of $0.34mm$. This difference, mainly due to unaddressed morphological defects, validates the good results obtained with the DIC measurement and highlights the advantages of the proposed method.

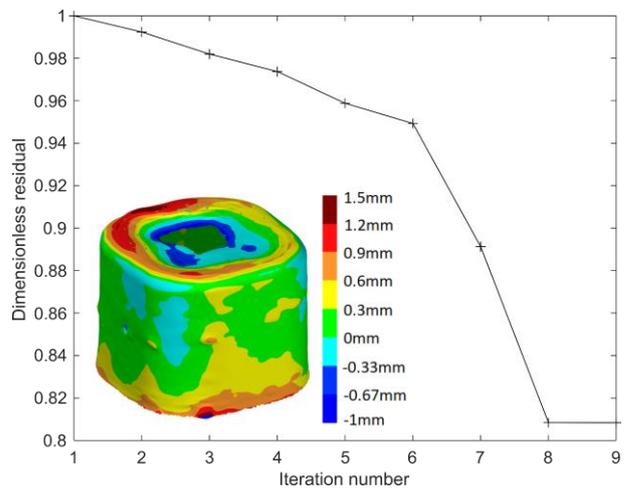

**Figure 3.** Convergence of the residual and the difference map between the optimized SE and the actual geometry.